\documentclass{PoS}

\usepackage{graphics}
\usepackage{graphicx}
\usepackage{amsmath}
\usepackage{cite}
\def\be{\begin{equation}}
\def\ee{\end{equation}}
\def\bea{\begin{eqnarray}}
\def\eea{\end{eqnarray}}
\def\eps{\epsilon}
\def\nnb{\nonumber}
\def\eps{\epsilon}
\def\dps{\displaystyle}

\def\cG{  {\cal G}  }
\def\cN{  {\cal N}  }
\def\cO{  {\cal O}  }
\def\Tr{  {\rm Tr}  }

\def\cA {  {\cal A}  }

\title{The Sudakov form factor to three loops in {$\mathbf{\mathcal{N}=4}$} super Yang-Mills}

\ShortTitle{The Sudakov form factor to three loops in $\mathcal{N}=4$ super Yang-Mills}

\author{\speaker{Tobias Huber}%
	 \\
        Theoretische Physik 1, Naturwissenschaftlich-Technische Fakult\"at,
\\ Universit\"at Siegen, Walter-Flex-Stra{\ss}e 3, D-57068 Siegen, Germany\\
        E-mail: \email{huber@tp1.physik.uni-siegen.de}}

\abstract{We review the results for the Sudakov form factor in $\cN=4$ super Yang-Mills theory up to
the three-loop level. At each loop order, the form factor is expressed as a linear combination of
only a handful scalar integrals, with small integer coefficients. Working in dimensional
regularisation, the expansion coefficients of each integral exhibit homogeneous transcendentality in
the Riemann $\zeta$-function. We find that the logarithm of the form factor reproduces the correct values of the cusp and
collinear anomalous dimensions. Moreover, the form factor in $\cN=4$ super Yang-Mills can be related
to the leading transcendentality pieces of the QCD quark and gluon form factor.
Finally, we comment briefly on the ultraviolet properties of the $\cN=4$ form factor in $D>4$ dimensions.}

\FullConference{
``Loops and Legs in Quantum Field Theory'' 11th DESY Workshop on Elementary Particle Physics \\
		 April 15-20, 2012\\
		 Wernigerode, Germany}

\begin{document}

\section{Introduction and definition}\label{sec:intro}

In recent years the investigation of scattering amplitudes in gauge theories --
in particular $\cN =4$ super Yang-Mills (SYM) theory --
has experienced tremendous progress, and revealed a lot of insight into their 
structure, see~\cite{review} for a review.

Quantities closely related to scattering amplitudes are form factors.
For example, planar amplitudes can be factorised into an infrared divergent part, given by a product of
form factors, and an infrared finite remainder~\cite{Bern:2005iz}.
The relation to form factors makes it possible to give an operator definition of the latter.
In addition, one observes that both scattering amplitudes and form factors
have uniform degree of transcendentality in their loop and/or $\eps$-expansion.

For both, the planar four-particle amplitude and the form factor, the general form of the 
result is known in principle. For the former, this is due to dual conformal symmetry,
for the latter it is due to the exponentiation of infrared divergences.
However, it is a non-trivial task to obtain these a priori known results
from an explicit linear combination of loop integrals. The final
result, however, is simple and suggests that there should be more structure hidden in
the loop integral expressions. Hence by studying them further one might
gain insights into better ways of evaluating them.

Despite the apparent simpler structure of form factors compared to scattering amplitudes
(the former have a trivial scale dependence), less is known
about the loop expansion of form factors in $\cN=4$ SYM than about scattering amplitudes.
For example, the calculation of the planar four-point amplitude has been carried out to
the four-loop order, see e.g.~\cite{Bern:2010tq}. On the other hand, the Sudakov
(or scalar) form factor in $\cN =4$ SYM has long been known only to two loops owing to a calculation by
van~Neerven~\cite{vanNeerven:1985ja}, and has only recently been extended to one higher loop~\cite{Gehrmann:2011xn}.

Although generalisations of the Sudakov form factor to the case of more external on-shell legs
and different composite operators have been discussed recently~\cite{Brandhuber:2010ad,Bork:2010wf}, 
we will restrict ourselves in the present article to the perturbative expansion of the Sudakov form factor discussed
in~\cite{vanNeerven:1985ja,Gehrmann:2011xn}.

We start by introducing the operator
\be\label{eq:composite}
\cO = \Tr(\phi_{12} \phi_{12}) \,,
\ee 
where the scalar fields $\phi_{AB}$ are in the representation $\bf 6$ of $SU(4)$,
and $\phi_{AB} = \phi_{AB}^{a} T_{a}$, with $T_{a}$ being
the generators of $SU(N)$ in the fundamental representation.
The operator $\cO$ is a colour singlet and has zero anomalous dimension.
In terms of $\cO$ the form factor is given by
\be
\mathcal{F}_{\cal S} = \langle \phi^{a}_{34}(p_1) \phi^{b}_{34}(p_2)\, \cO \rangle 
\equiv \Tr(T^{a} T^{b}) F_S \, .
\ee
The states $\phi^{a}_{34}(p_1)$ and $\phi^{b}_{34}(p_2)$ are in the adjoint representation, and the
outgoing momenta $p_1$ and $p_2$ are massless and on-shell, i.e.\ $p_1^2=p_2^2=0$, and $q^2\equiv
(p_1+p_2)^2$. In order to regularise IR divergences associated with the on-shell legs we work in
dimensional regularisation with $D=4-2\eps$. In order to facilitate the presentation of the results
in sections~\ref{sec:ff3l} and~\ref{sec:logff} we introduce two more quantities, the first one being
the dimensionless variable $\dps x = \mu^2/(-q^2-i\eta)\, ,$ with infinitesimal $\eta>0$.
The second quantity is the 't~Hooft coupling
$\dps a = (g^2 \, N)/(8\pi^2) \, (4\pi)^\eps \, e^{-\eps\gamma_E}\,,$
where $g$ is the gauge coupling of $\cN = 4$~SYM, $N$ is the number of colours, and $\gamma_{E}\approx
0.5772$ is the Euler-Mascheroni constant. The loop-expansion of the form factor now assumes the following
form,
\bea\label{eq:ffexpansion1}
\dps {F}_S &=& 1+ a \, x^{\eps} \, {F}_S^{(1)}
+ a^2 \, x^{2\eps} \, {F}_S^{(2)}
+ a^3 \, x^{3\eps} \, {F}_S^{(3)}+{\cal{O}}(a^4)\; .
\eea
The superscripts denote the loop-order, and we normalised the tree-level contribution to unity.

Up to the three-loop level, the $L$-loop form factor ${F}_S^{(L)}$ is strictly proportional to $N^L$,
i.e.\ there is only the leading-in-colour contribution. This changes at four loops since the
quartic Casimir $(d_{abcd})^2$ can appear. Whether or not the latter will actually be present at four
loops is another very interesting related question, and has to do with the colour 
dependence of infrared divergences in gauge theories, see e.g.\ \cite{Becher:2009qa} and references
therein.

\section{Derivation of the form factor from unitarity cuts}\label{sec:ff3luni}

{}
\FIGURE[t]{
\includegraphics[width=0.95\textwidth]{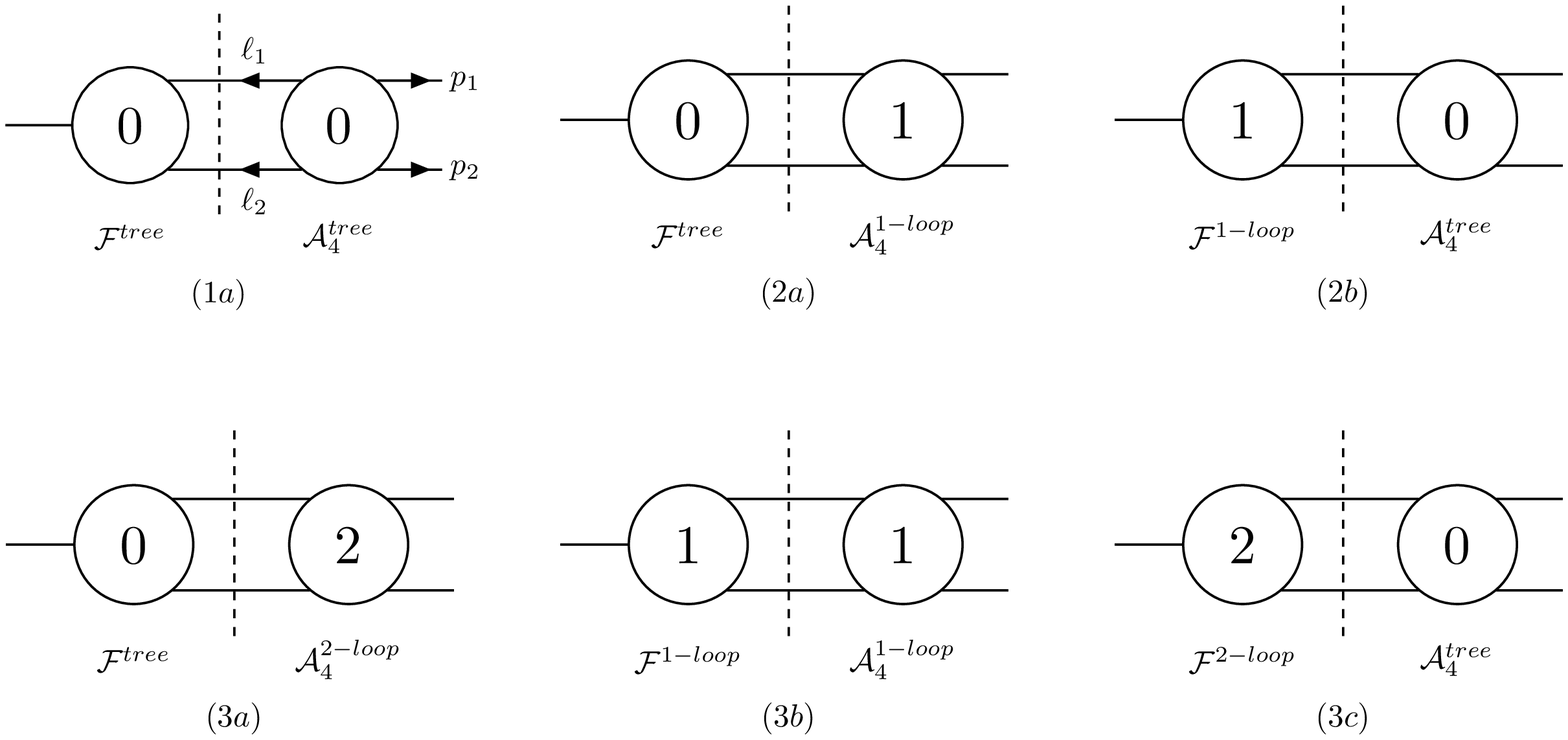}
\caption{Two-particle cuts up to three loops. The numbers inside the circles indicate the respective number of loops in
the form factors and four-particle scattering amplitudes.}
\label{fig:2pcuts}}

We will use the method of unitarity cuts~\cite{Bern:1994zx,Britto:2004nc} to derive an expression for the Sudakov form
factor in $\mathcal{N}=4$ SYM in terms of scalar loop integrals. We will apply two-particle cuts, as well as generalised
cuts. The two-particle cuts are displayed schematically in Fig.~\ref{fig:2pcuts}. At a given loop order $L \ge 1$ one has
to consider all contributions from cuts of the $m$-loop form factor with the $(L-1-m)$-loop four-particle scattering
amplitude, with $m=0,\ldots,L-1$. The respective values are shown inside the circles in Fig.~\ref{fig:2pcuts}.

Let us derive the one-loop result explicitly. We follow the notations for unitarity cuts of ref.~\cite{hep-ph/9702424}. 
We have to compute the two-particle cut (1a) shown in Fig.~\ref{fig:2pcuts}. It is given by
\be\label{eq:cut1loop}
{\cal F}_{\cal S}^{1-loop} \Bigr|_{{\rm cut (1a)}} =  \int \sum_{P_1 , P_2 }  \frac{d^{D}k}{(2\pi)^{D}} \, \frac{i}{\ell_2^2} \, {\cal F}_{S}^{tree}(-\ell_1 , -\ell_2 ) \,  \frac{i}{\ell_1^2} \, \mathcal{A}_{4}^{tree}(\ell_2 , \ell_1, p_1 , p_2 )  \Bigr|_{\ell_1^2 = \ell_2^2=0} \,,
\ee
where $\ell_1$ and $\ell_2$ are the momenta of the cut legs, and the sum runs over all possible particles across the cut.
The four-particle tree ampliutde $\mathcal{A}_{4}^{tree}(\ell_2 , \ell_1, p_1 , p_2 )$ is given by
\be\label{app:4pttree}
\cA_{4}^{tree} = g^2 \mu^{2 \eps} \sum_{ \sigma \in S_{4}/Z_{4}} \Tr(T^{a_{\sigma(1)}}  T^{a_{\sigma(2)} }T^{a_{\sigma(3)} }T^{a_{\sigma(4)}} ) A_{4;1;1}^{tree}(\sigma(1), \sigma(2),\sigma(3),\sigma(4)) \,,
\ee
with the `partial amplitudes'
$A_{4;1,1}^{tree}(\phi_{12}(1),\phi_{12}(2),\phi_{34}(3),\phi_{34}(4))  = - i\, s_{12}/s_{23} \,$.
The tree-level form factor is simply given by
\be
{\cal F}_{\cal S}^{tree}(-\ell_1 , -\ell_2 ) = \Tr(T^{a} T^{b}) \,.
\ee
With our choice of external states, only scalars can appear as intermediate particles, and we do not need the spinor
helicity formalism. With this, Eq.~(\ref{eq:cut1loop}) becomes
\bea\label{eq:cut1loopb}
{\cal F}_{\cal S}^{1-loop} \Bigr|_{{\rm cut\,(1a)}} &=& 
 -2\,  g^2 \mu^{2 \eps}\, N \,  q^2 \, \Tr(T^{a} T^{b} )\, 
\int  \frac{d^{D}k}{i (2\pi)^{D}} \frac{1}{k^2 (k+p_{1})^2 (k-p_{2})^2} \,
 \Bigr|_{{\rm cut\,(1a)}} \nnb \\
 &=&-2\,  g^2 \mu^{2 \eps}\, N \,  q^2 \, \Tr(T^{a} T^{b} )\, D_1 \,
 \Bigr|_{{\rm cut\,(1a)}}\,,
\eea
where we have identified the cut of the one-loop form factor with
the cut of the one-loop triangle integral $D_{1}$, see Fig.~\ref{fig:diagsDE}.
It turns out that this result is exact, i.e.\ that we can remove the ``cut (1a)'' in Eq.~(\ref{eq:cut1loopb})
and get
\be
{F}_{S}^{1-loop} = g^2  N \mu^{2 \eps} (-q^2) 2 D_1  \,.
\ee

At two loops, following analogous steps, the result for the form factor is given by~\cite{vanNeerven:1985ja},
\be\label{eq:vanNeerven}
{F}_{S}^{2-loop} = g^4 N^2 
\mu^{4 \eps} (-q^2)^2 \big\lbrack 4 E_1 + E_2 \big\rbrack  \,,
\ee
where the diagrams $E_1$ and $E_2$ are also shown in Fig.~\ref{fig:diagsDE}.
The unitarity cut (2b) of Fig.~\ref{fig:2pcuts} detects only the presence of the planar integral $E_1$.
The unitarity cut (2a) of Fig.~\ref{fig:2pcuts} reveals -- besides $E_1$ -- the non-planar
integral $E_2$. The appearance of the latter stems from the fact that 
we have to use the full one-loop four-point amplitude
\bea
\label{app:4pt1loop}
\cA_{4}^{1-loop} &=& g^4 \mu^{4 \eps}  \sum_{ \sigma \in S_{4}/Z_{4}}   N \, \Tr(T^{a_{\sigma(1)}}  T^{a_{\sigma(2)} }T^{a_{\sigma(3)} }T^{a_{\sigma(4)}} ) A_{4;1,1}^{1-loop}(\sigma(1), \sigma(2),\sigma(3),\sigma(4))  \nonumber  \\
&& \hspace{-1.3cm}+ g^4 \mu^{4 \eps} \sum_{ \sigma \in S_{4}/Z_{2}^3}   \Tr(T^{a_{\sigma(1)}}  T^{a_{\sigma(2)} }) \Tr( T^{a_{\sigma(3)} }T^{a_{\sigma(4)}} ) A_{4;1,3}^{1-loop}(\sigma(1), \sigma(2),\sigma(3),\sigma(4)) \,,
\eea
which in addition to single trace terms also contains double trace terms. The latter are subleading 
in the number of colours $N$. However, the colour algebra gives rise to another
factor of $N$ for those terms, so that they contribute to the form factor at the leading colour, just like
the single trace terms.

Finally, at three loops the two-particle cuts are given by cuts (3a)~--~(3c) of Fig.~\ref{fig:2pcuts}. One finds for their
total contribution
\be\label{result-2pcuts}
{ F}_{S}^{3-loop} \Bigr|_{{\rm 2-part.\; cut}} = 
g^6\,\mu^{6\epsilon}\,N^3 \, (-q^2)^2 \big\lbrack 
8\, (-q^2) \,  F_{1} - 2\,  F_{2} + 4\,  F_{3} + 4  \, F_{4} -4\,  F_{5} - 4 \, F_{6} - 4\,  F_{8} \big\rbrack \Bigr|_{{\rm 2-part.\; cut}} \,.
\ee
The integrals $F_i$ are given in Fig.~\ref{fig:diagsF}. It is remarkable that the coefficients of all integrals are small
integer numbers. In order to detect also integrals not having any two-particle cuts we study generalised cuts, where we
cut all or all but one propagator. This serves as a cross-check on the results already obtained above and detects further
integrals such as $F_9$. The total result at three loops then assumes the form
\bea\label{eq:3lfirst}
{F}_S^{3-loop} &=&   g^6 \, \mu^{6\eps}  \, N^3 \, (-q^2)^2 \,
\big\lbrack 8 \, (-q^2) \, F_1 - 2 \, F_2 + 4 \, F_3 + 4 \, F_4 - 4 \,
F_5 - 4 \, F_6 - 4 \, F_8 + 2 \, F_9 \big\rbrack \,. %\nonumber \\ 
\eea

%%%%%%%%%%%%%%%%%%%%%%%%%%%%%%%%%%%%%%%%%%%%%%%%%%%%%%%%%%%%%%%%%%%%%%%%%%%%%%%%%%%
%%%%%%%%%%%%%%%%%%%%%%%%%%%%%%%%%%%%%%%%%%%%%%%%%%%%%%%%%%%%%%%%%%%%%%%%%%%%%%%%%%%

\section{Final result for the form factor up to three loops}\label{sec:ff3l}

{}
\FIGURE[t]{
\includegraphics[width=0.76\textwidth]{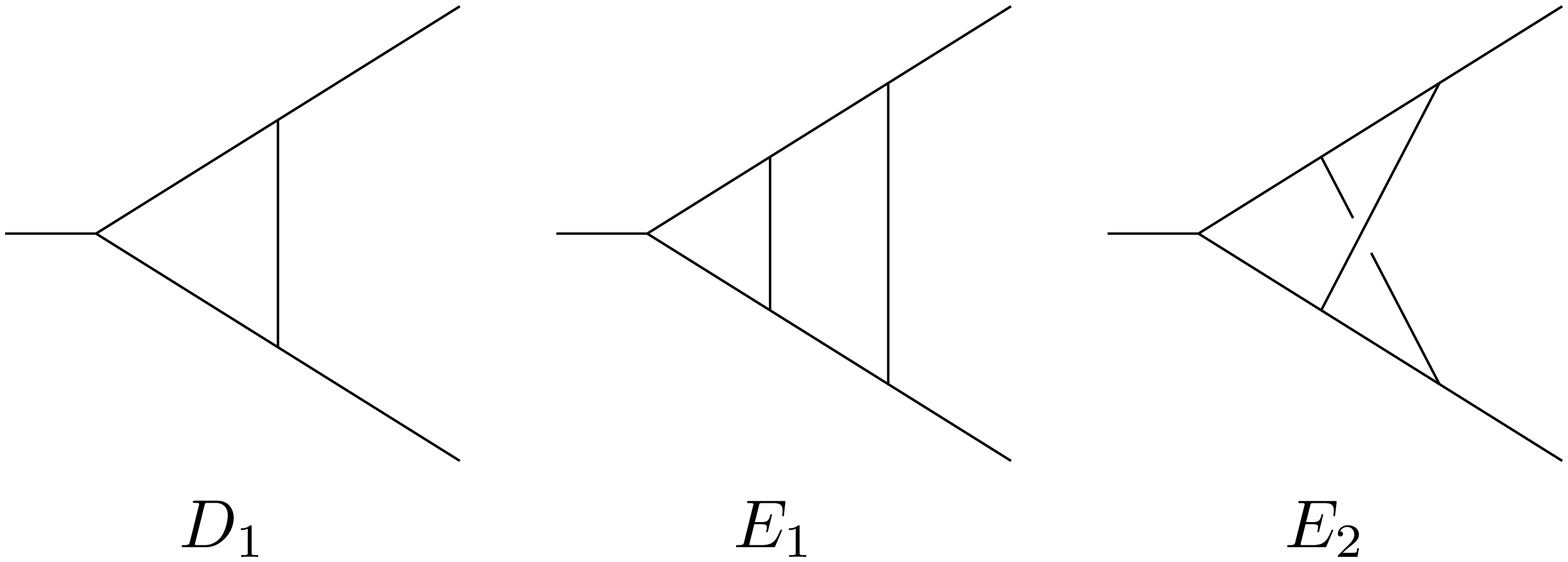}
\caption{Diagrams that contribute to the one-loop and
two-loop form factor in $\mathcal{N}=4$ SYM. All internal
lines are massless.}
\label{fig:diagsDE}}

Using unitarity cut methods described in the previous section we obtain the following result for the
$\cN=4$ SYM form factor up to three loops~\cite{Gehrmann:2011xn}.
{\allowdisplaybreaks
\bea
{F}_S &=& 1+ g^2 \, N \, \mu^{2\eps} \cdot (-q^2)\cdot 2\, D_1
+ g^4 \, N^2 \, \mu^{4\eps} \cdot (-q^2)^2 \cdot
\left[4 \, E_1 + E_2\right]  \nnb \\
&& + \, g^6 \, N^3 \, \mu^{6\eps} \cdot (-q^2)^2 \cdot
\left[8 \, (-q^2) \, F_1 - 2 \, F_2 + 4 \, F_3 + 4 \, F_4 - 4 \, F_5 - 4 \, F_6 - 4 \, F_8 + 2 \, F_9\right] \nnb \\
&&+ \, {\cal{O}}(g^8) \; . \label{eq:3lut}
\eea
}
All diagrams are shown in Figs.~\ref{fig:diagsDE} and~\ref{fig:diagsF}. It is remarkable that the
form factor up to three loops is given by a small number of scalar loop integrals, each having a
small integer coefficient. Working in dimensional regularisation with $D=4-2\eps$, the Laurent-series
expansions of all diagrams are known from the calculation of the QCD quark and gluon form
factor~\cite{Gehrmann:2005pd,masterC,BCSSS,arXiv:1001.2887,Gehrmann:2010ue,Lee:2010ik,arXiv:1010.4478}.
They yield for the Sudakov form factor in $\mathcal{N}=4$ SYM
{\allowdisplaybreaks
\bea
{F}_S^{(1)} &=& -\frac{1}{\eps^2}+\frac{\pi ^2}{12}+\frac{7\zeta_3}{3} \, \eps
+ \frac{47 \pi ^4}{1440} \, \eps^2 +\eps^3 \left(\frac{31
\zeta_5}{5}-\frac{7 \pi ^2\zeta_3}{36}\right) +\eps^4 \left(\frac{949
\pi ^6}{120960}-\frac{49\zeta_3^2}{18}\right)\nnb \\
&& +\eps^5\left(-\frac{329
\pi ^4 \zeta_3}{4320}-\frac{31 \pi ^2 \zeta_5}{60}+\frac{127
\zeta_7}{7}\right) +\eps^6 \left(\frac{49 \pi ^2
\zeta_3^2}{216}-\frac{217 \zeta_3 \zeta_5}{15}+\frac{18593 \pi
^8}{9676800}\right) \nnb \\
&& + {\cal{O}}(\eps^7) \; , \label{eq:1lff} \\
&& \nnb \\
{F}_S^{(2)} &=&+\frac{1}{2 \eps^4}-\frac{\pi ^2}{24\eps^2}-\frac{25 \zeta_3}{12
\eps}-\frac{7 \pi^4}{240}
+\eps \left(\frac{23 \pi ^2 \zeta_3}{72}+\frac{71 \zeta_5}{20}\right)
+\eps^2 \left(\frac{901 \zeta_3^2}{36}+\frac{257 \pi
^6}{6720}\right)\nnb \\
&&+\eps^3 \left(\frac{1291 \pi
   ^4 \zeta_3}{1440}-\frac{313 \pi ^2 \zeta_5}{120}+\frac{3169
   \zeta_7}{14}\right) \nnb \\
&&+\eps^4 \left(-66 \zeta_{5,3}+\frac{845 \zeta_3
\zeta_5}{6}-\frac{1547 \pi ^2 \zeta_3^2}{216}+\frac{50419 \pi
   ^8}{518400}\right)+ {\cal{O}}(\eps^5) \; , \label{eq:2lff} \\
{F}_S^{(3)} &=& -\frac{1}{6 \eps^6}+\frac{11 \zeta_3}{12 \eps^3}+\frac{247 \pi ^4}{25920 \eps^2}
+\frac{1}{\eps}\left(-\frac{85 \pi ^2 \zeta_3}{432}-\frac{439 \zeta_5}{60}\right) \nnb \\
&& -\frac{883 \zeta_3^2}{36}-\frac{22523 \pi ^6}{466560}
+\eps \left(-\frac{47803 \pi ^4 \zeta_3}{51840}+\frac{2449 \pi ^2 \zeta_5}{432}-\frac{385579 \zeta_7}{1008}\right) \nnb \\
&& + \eps^2 \left(\frac{1549}{45} \zeta_{5,3}-\frac{22499 \zeta_3 \zeta_5}{30}+\frac{496 \pi ^2 \zeta_3^2}{27}
-\frac{1183759981 \pi ^8}{7838208000}\right) + {\cal{O}}(\eps^3) \; . \label{eq:3lff}
\eea}
{}
\FIGURE[t]{
\includegraphics[width=0.98\textwidth]{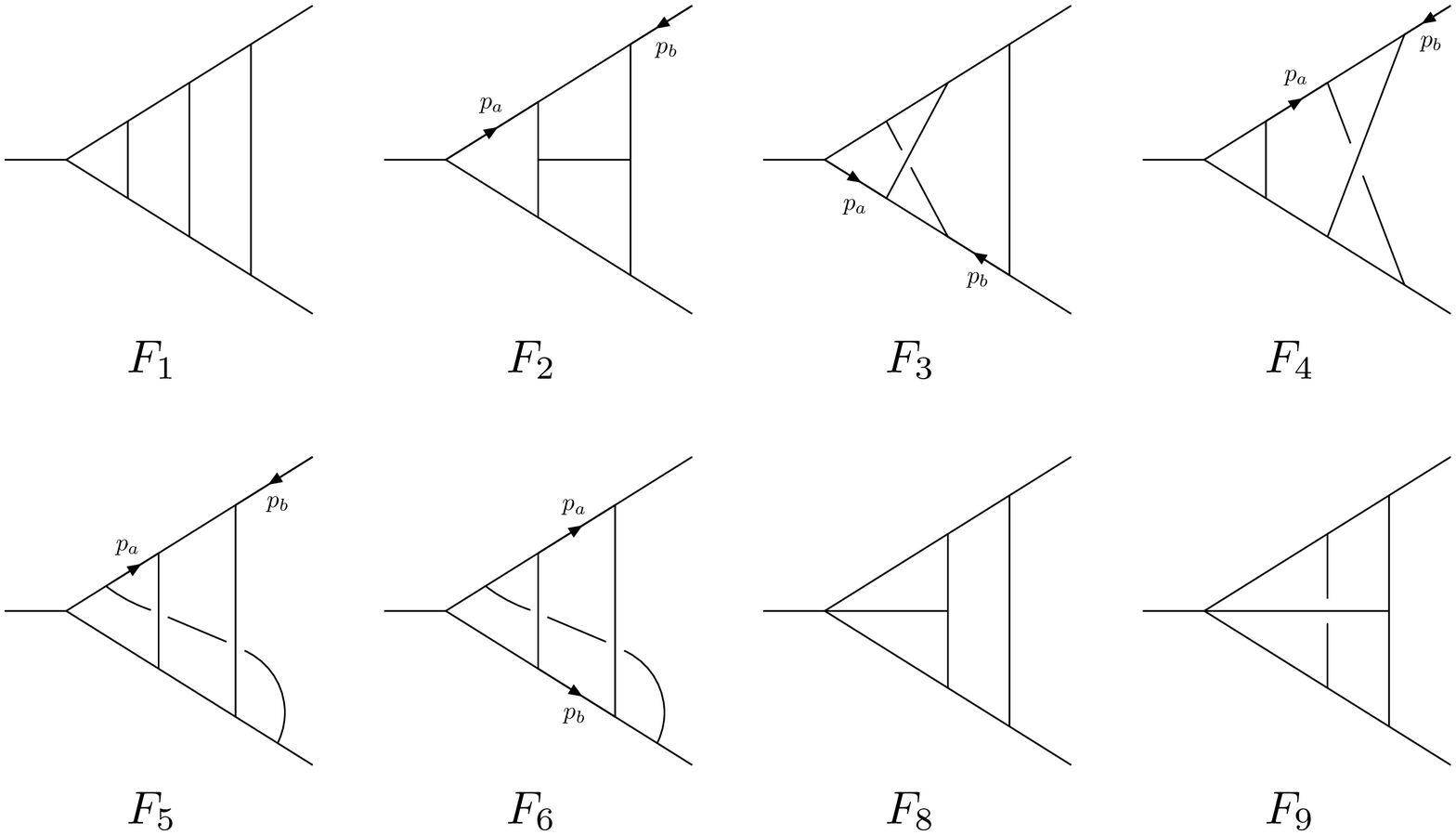}
\caption{Diagrams that contribute to the three-loop form factor
in $\mathcal{N}=4$ SYM. All internal lines are massless.
$p_a$ and $p_b$ on arrow lines denote an irreducible scalar product
$(p_a+p_b)^2$ in the numerator.}
\label{fig:diagsF}}
The coefficients of the $\eps$-expansions are of increasing transcendentality (or weight) in the
Riemann $\zeta$-function\footnote{One assigns to $\pi^i$ the weight $i$ and to $\zeta_k$ the weight
$k$. Their product has weight $i+k$.}. One recognizes that each coefficient in the above formulas has
homogeneous weight; a property that does not only hold true for the final result, but for each of the
diagrams in Eq.~(\ref{eq:3lut}) contributing to it.
We also remark that in order to obtain all finite pieces of the logarithm of the form factor (see
section~\ref{sec:logff}) we need the $\eps$-expansion through terms of transcendental weight six. We
emphasize that our expressions contain two more orders in $\eps$ and therefore contain already all
information required for exponentiation at four loops.

Let us elaborate here on yet another very interesting observation, namely the leading
transcendentality principle~\cite{hep-th/0404092}. To this end, let us specify the QCD quark and
gluon form factor -- which do {\emph{not}} have the homogeneous-weight property -- to a supersymmetric
Yang-Mills theory with a bosonic and fermionic degree of freedom in the same colour representation.
This is achieved by setting $C_A = C_F = 2 \, T_F$ and $n_f=1$ in the QCD
result~\cite{BCSSS}. We find that with this adjustment the leading (i.e.\ highest)
transcendentality pieces of the quark and gluon form factor become
equal, and moreover coincide with the Sudakov form factor in $\cN=4$~SYM presented here.
This equality holds true at one, two, and three loops and in all coefficients up to
transcendental weight eight, and it serves as an important check of our result.

%%%%%%%%%%%%%%%%%%%%%%%%%%%%%%%%%%%%%%%%%%%%%%%%%%%%%%%%%%%%
%%%%%%%%%%%%%%%%%%%%%%%%%%%%%%%%%%%%%%%%%%%%%%%%%%%%%%%%%%%%

\section{Logarithm of the form factor}\label{sec:logff}
The logarithm of the form factor is given by
{\allowdisplaybreaks
\bea
\dps \ln\left({F}_S\right) &=& 
\ln\left(1+ a \, x^{\eps} \, {F}_S^{(1)}
+ a^2 \, x^{2\eps} \, {F}_S^{(2)}
+ a^3 \, x^{3\eps} \, {F}_S^{(3)}+{\cal{O}}(a^4)\right) \nnb\\
&=& a \, x^{\eps} \, {F}_S^{(1)}
+ a^2 \, x^{2\eps} \left[{F}_S^{(2)}-\frac{1}{2} \left({F}_S^{(1)}\right)^2\right]
+ a^3 \, x^{3\eps} \left[{F}_S^{(3)}-{F}_S^{(1)}{F}_S^{(2)}
+\frac{1}{3}\left({F}_S^{(1)}\right)^3\right] \nnb\\
&&+{\cal{O}}(a^4) \; .
\eea
}
Plugging in the results from Eqs.~(\ref{eq:1lff})~--~(\ref{eq:3lff}) we verify the cancellation of
all poles higher than $1/\eps^2$, as expected from exponentiation of infrared divergences. The
logarithm of the form factor therefore has the generic structure~\cite{Magnea:1990zb}
\be
\ln\left({F}_S\right) =
\sum\limits_{L=1}^{\infty} a^L \, x^{L\eps} \,
\left[ 
-\frac{\gamma^{(L)}}{ 4 (L \eps)^2} 
-\frac{\cG_0^{(L)}} {2 L \eps} 
\right] + {\cal{O}}(\eps^0) \; ,
\ee
and we confirm up to $L=3$ the $L$-loop cusp $\gamma^{(L)}$    
and collinear $\cG_0^{(L)} $ 
anomalous dimensions~\cite{Korchemskaya:1992je}
\bea
\label{eq:gammacusp}
\gamma(a) 
&=& 
\sum\limits_{L=1}^\infty a^L  
\gamma^{(L)}
= 
4a - 4\zeta_2 a^2 + 22 \zeta_4 a^3 + {\cal{O}}(a^4) \; ,
\\
\label{eq:gammaG}
\cG_0(a)
&=& 
\sum\limits_{L=1}^\infty  a^L  \cG_0^{(L)} =
    -  \zeta_3 a^2 + \left(4 \zeta_5 + \frac{10}{3} \zeta_2 \zeta_3
    \right)  a^3 + {\cal{O}}(a^4) \; .
\eea

%%%%%%%%%%%%%%%%%%%%%%%%%%%%%%%%%%%%%%%%%%%%%%%%%%%%%%%%%%%%
%%%%%%%%%%%%%%%%%%%%%%%%%%%%%%%%%%%%%%%%%%%%%%%%%%%%%%%%%%%%

\section{Ultraviolet divergences in higher dimensions}\label{sec:uv}

The Sudakov form factor is ultraviolet (UV) finite in $D=4$ dimensions. One can now study
the form factor as a function of the number $D$ of space-time dimensions and investigate
at which $D$ it first develops UV divergences. This particular $D$ is called ``critical
dimension'' and depends on the number of loops. Hence we denote it by $D_{c}(L)$.
The knowledge of $D_c$ at a given loop order is
useful since it can allow for a cross-check of computations, or constrain the types of loop
integrals that can appear (or, even more important, that {\emph{cannot}} appear).
There is a  bound on $D_c$ based on power counting for supergraphs and the background field 
method which reads~\cite{Grisaru:1982zh,Marcus:1984ei},
\be \label{susyboundUV}
D_{c}(L) \; \ge \; 4 + \frac{2 (\cN-1)}{L} \; = \; 4 + \frac{4}{L} \,, \qquad L>1\, .
\ee
The formula is valid for $L>1$ only. For $D<D_{c}$ the theory is UV finite. We plugged in
$\cN=3$ in~(\ref{susyboundUV}) since here $\cN$  denotes on the number of supersymmetries that
can be realized off-shell.

We will now investigate whether the lower bound~(\ref{susyboundUV}) for $D_{c}$ is saturated, or if
the formula gives a bound that is too conservative. There is no statement
from Eq.~(\ref{susyboundUV}) for the one-loop case, but one can
easily see from Fig.~\ref{fig:diagsDE} that $D_{c}(L=1) = 6$. From the same
Figure, one can see that also at two-loops we have $D_{c}(L=2) = 6$, which follows from
na\"ive power counting. Hence at two-loops the
bound (\ref{susyboundUV}) is indeed saturated.
At three loops,  Eq.~(\ref{susyboundUV}) becomes $D_{c} \ge 16/3$.
We will now investigate if we have ${D_{c}(L=3) = 16/3}$ or if the form
factor at three loops is better behaved in the UV than expected from~(\ref{susyboundUV}).
To this end we take the UV limit of the three-loop term of Eq.~(\ref{eq:3lut}) by
giving all propagators (and also all numerators) a
common mass $m$ and by nullifying the external momenta. This is
possible since there are no sub-divergences in $D=16/3$. In this limit
we get~\cite{Gehrmann:2011xn}
\be\label{eq:FFUV}
{F}_S^{3-loop} \propto
(-q^2) \, \left[ 8 \, F_1 + 2 \, F_3^* + 2 \, F_4^*\right] - 2 \, F_2 + 4 \, F_5^* - 2 \, F_9 \,.
\ee
where the asterisk on $F_3$ and $F_4$ indicates the respective integral with unit numerator.
$F_5^*$ is obtained from $F_5$ by replacing in the numerator
$(p_a^{F_5}+p_b^{F_5})^2 \longrightarrow (p_a^{F_6}+p_b^{F_6}-p_a^{F_5}+p_b^{F_5})^2$.
The first three integrals are finite by na\"ive power counting, and the
last three integrals become equal in the aforementioned UV limit, and cancel due to their pre-factors.
This renders the three-loop form factor finite in $D=16/3$ dimensions. It is therefore better
behaved in the UV than suggested by Eq.~(\ref{susyboundUV}).

The next value of $D$ where the form factor can -- and indeed does -- develop UV divergences is
$D_c(L=3)=6$. We have therefore found $D_c(L)=6$ for~$L=1$,~$2$,~$3$. We now take a closer look at the
UV properties of the form factor in six dimensions. Specifying  $D=6-2\eps$ and taking the aforementioned
UV limit we find that the leading UV pole at $L$ loops is $1/\eps^L$. Moreover, the leading pole is always
produced by the $L$-loop planar ladder diagram.  All other diagrams start at most at a subleading
pole in $\eps$. When considering $\log(F_{S})$ in the UV limit all higher poles cancel and there are only
simple $1/\eps$ poles up to three loops.

An equation similar to~(\ref{susyboundUV}) holds also for scattering amplitudes in the UV limit. In this case one even
finds the stronger bound $D_{c}(L) \ge 4 + {6}/{L}$, which is saturated at two and three loops~\cite{Bern:2010tq}. At one
loop one finds $D_{c}(L=1)=8$ for the four-particle scattering amplitude. So despite the fact that the form factor is
better behaved in the UV than expected, four-particle scattering amplitudes are even better behaved in the UV than the
form factor. One reason for this is the fact that amplitudes, at least in the planar limit, 
are dual conformal invariant, whereas form factors are not. Another reason is the fact that in $D=6$ the operator $\cO$
in~(\ref{eq:composite}) has the counterterm $g^2 \, \square\, {\rm tr}\, ( \phi^2 )$, and other operators having the same
quantum numbers; and operator mixing can occur at one loop.

\section{Conclusion}\label{sec:conc}

We presented the results for the Sudakov form factor in $\cN=4$ super Yang-Mills theory up to
the three-loop level. We employed the unitarity-based method to derive the answer in terms of
both, planar and non-planar loop integrals. At each loop order, the form factor is expressed as
a linear combination of only a handful scalar integrals, with small integer coefficients.
We evaluated the form factor in dimensional regularisation to $\cO(\epsilon^{8-2 L})$ ($L$ is the number of loops) and
found that the expansion coefficients of each integral exhibit homogeneous transcendentality in the Riemann
$\zeta$-function. Moreover, we verified the exponentiation of infrared divergences,
and reproduced the correct values of the cusp and collinear anomalous dimensions.

In addition, we observed that the heuristic leading transcendentality principle that
relates anomalous dimensions in QCD with those in $\cN=4$ SYM 
also holds for the form factor. We verified this principle to three loops, and
through to terms of transcendentality eight.

Finally, we studied the UV behaviour of the form factor in higher dimensions, and found that 
the critical dimension ist given by $D_c(L)=6$ up to three loops.
This means that the three-loop result is better behaved in the UV than suggested by Eq.~(\ref{susyboundUV}). In
particular, it is finite in $D=16/3$ dimensions.

An interesting further direction of the present calculation would be its extension to
four loops, since it would allow to get insight into the non-planar colour structure. Whether the
anomalous dimension associated with the quartic Casimir $(d_{abcd})^2$ vanishes is a hot topic
and has to do with the general question of colour dependence of infrared divergences in gauge
theories~\cite{Becher:2009qa}.

%%%%%%%%%%%%%%%%%%%%%%%%%%%%%%%%%%%%%%%%%%%%%%%%%%%%%%%%%%%%
%%%%%%%%%%%%%%%%%%%%%%%%%%%%%%%%%%%%%%%%%%%%%%%%%%%%%%%%%%%%

\section*{Acknowledgements}
I would like to thank the organizers of ``Loops~and~Legs~2012'' for creating a pleasant and inspiring
atmosphere. Special thanks goes to Thomas~Gehrmann and Johannes~Henn for a fruitful collaboration.
The work of the author was supported by the Helmholtz-Alliance ``Physics at the Terascale''.


\begin{thebibliography}{99}

%\cite{review}
\bibitem{review}
R.~Roiban, M.~Spradlin and A.~Volovich (ed),
%``Scattering amplitudes in gauge theories: progress and outlook,''
J. Phys. A: Math. Theor. {\bf 44} 450301 (2011)

%\cite{Bern:2005iz}
\bibitem{Bern:2005iz}
  Z.~Bern, L.~J.~Dixon and V.~A.~Smirnov,
  %``Iteration of planar amplitudes in maximally supersymmetric Yang-Mills theory at three loops and beyond,''
  Phys.\ Rev.\ D\ {\bf 72} (2005) 085001
  [hep-th/0505205].
  %%CITATION = PHRVA,D72,085001;%%

%\cite{Bern:2010tq}
\bibitem{Bern:2010tq}
  Z.~Bern, J.~J.~M.~Carrasco, L.~J.~Dixon, H.~Johansson, R.~Roiban,
  %``The Complete Four-Loop Four-Point Amplitude in N=4 Super-Yang-Mills Theory,''
  Phys.\ Rev.\  {\bf D82 } (2010)  125040.
  [arXiv:1008.3327 [hep-th]].

%\cite{vanNeerven:1985ja}
\bibitem{vanNeerven:1985ja}
  W.~L.~van Neerven,
  %``Infrared Behavior Of On-shell Form-factors In A N=4 Supersymmetric Yang-mills Field Theory,''
  Z.\ Phys.\ C\ {\bf 30} (1986) 595.
  %%CITATION = ZEPYA,C30,595;%%

%\cite{Gehrmann:2011xn}
\bibitem{Gehrmann:2011xn}
  T.~Gehrmann, J.~M.~Henn and T.~Huber,
  %``The three-loop form factor in N=4 super Yang-Mills,''
  JHEP {\bf 1203} (2012) 101
  [arXiv:1112.4524 [hep-th]].
  %%CITATION = ARXIV:1112.4524;%%

%\cite{Brandhuber:2010ad}
\bibitem{Brandhuber:2010ad}
  A.~Brandhuber, B.~Spence, G.~Travaglini, G.~Yang,
  %``Form Factors in N=4 Super Yang-Mills and Periodic Wilson Loops,''
  JHEP {\bf 1101 } (2011)  134.
  [arXiv:1011.1899 [hep-th]].

%\cite{Bork:2010wf}
\bibitem{Bork:2010wf}
  L.~V.~Bork, D.~I.~Kazakov, G.~S.~Vartanov,
  %``On form factors in N=4 sym,''
  JHEP {\bf 1102 } (2011)  063.
  [arXiv:1011.2440 [hep-th]].

%\cite{Becher:2009qa}
\bibitem{Becher:2009qa}
  T.~Becher and M.~Neubert,
  %``On the Structure of Infrared Singularities of Gauge-Theory Amplitudes,''
  JHEP {\bf 0906} (2009) 081
  [arXiv:0903.1126 [hep-ph]].
  %%CITATION = ARXIV:0903.1126;%%

%\cite{Bern:1994zx}
\bibitem{Bern:1994zx}
  Z.~Bern, L.~J.~Dixon, D.~C.~Dunbar and D.~A.~Kosower,
  %``One loop n point gauge theory amplitudes, unitarity and collinear limits,''
  Nucl.\ Phys.\ B {\bf 425} (1994) 217
  [hep-ph/9403226].
  %%CITATION = HEP-PH/9403226;%%

%\cite{Britto:2004nc}
\bibitem{Britto:2004nc}
  R.~Britto, F.~Cachazo and B.~Feng,
  %``Generalized unitarity and one-loop amplitudes in N=4 super-Yang-Mills,''
  Nucl.\ Phys.\ B {\bf 725} (2005) 275
  [hep-th/0412103].
  %%CITATION = HEP-TH/0412103;%%

%\cite{hep-ph/9702424}
\bibitem{hep-ph/9702424}
  Z.~Bern, J.~S.~Rozowsky and B.~Yan,
  %``Two loop four gluon amplitudes in N=4 superYang-Mills,''
  Phys.\ Lett.\ B\ {\bf 401} (1997) 273
  [hep-ph/9702424].
  %%CITATION = PHLTA,B401,273;%%

%\cite{Gehrmann:2005pd}
\bibitem{Gehrmann:2005pd}
  T.~Gehrmann, T.~Huber, D.~Ma\^itre,
  %``Two-loop quark and gluon form-factors in dimensional regularisation,''
  Phys.\ Lett.\  {\bf B622 } (2005)  295-302.
  [hep-ph/0507061].

%\cite{Heinrich:2009be}
\bibitem{masterC}
  G.~Heinrich, T.~Huber, D.~A.~Kosower and V.~A.~Smirnov,
  %``Nine-Propagator Master Integrals for Massless Three-Loop Form Factors,''
  Phys.\ Lett.\  B {\bf 678} (2009) 359
  [arXiv:0902.3512 [hep-ph]].
  %%CITATION = PHLTA,B678,359;%%

%\cite{Baikov:2009bg}
\bibitem{BCSSS}
 P.~A.~Baikov, K.~G.~Chetyrkin, A.~V.~Smirnov, V.~A.~Smirnov and M.~Steinhauser,
  %``Quark and gluon form factors to three loops,''
  Phys.\ Rev.\ Lett.\  {\bf 102} (2009) 212002
  [arXiv:0902.3519 [hep-ph]].
  %%CITATION = PRLTA,102,212002;%%

%\cite{arXiv:1001.2887}
\bibitem{arXiv:1001.2887}
  R.N.~Lee, A.V.~Smirnov and V.A.~Smirnov,
  %``Analytic Results for Massless Three-Loop Form Factors,''
  JHEP {\bf 1004} (2010) 020  [arXiv:1001.2887 [hep-ph]].
  %%CITATION = ARXIV:1001.2887;%%

%\cite{Gehrmann:2010ue}
\bibitem{Gehrmann:2010ue}
  T.~Gehrmann, E.~W.~N.~Glover, T.~Huber, N.~Ikizlerli and C.~Studerus,
  %``Calculation of the quark and gluon form factors to three loops in QCD,''
  JHEP {\bf 1006} (2010) 094
  [arXiv:1004.3653 [hep-ph]].
  %%CITATION = JHEPA,1006,094;%%

%\cite{Lee:2010ik}
\bibitem{Lee:2010ik}
  R.~N.~Lee and V.~A.~Smirnov,
  %``Analytic Epsilon Expansions of Master Integrals Corresponding to Massless Three-Loop Form Factors and Three-Loop g-2 up to Four-Loop Transcendentality Weight,''
  JHEP\ {\bf 1102} (2011) 102
  [arXiv:1010.1334 [hep-ph]].
  %%CITATION = JHEPA,1102,102;%%

%\cite{arXiv:1010.4478}
\bibitem{arXiv:1010.4478}
  T.~Gehrmann, E.~W.~N.~Glover, T.~Huber, N.~Ikizlerli and C.~Studerus,
  %``The quark and gluon form factors to three loops in QCD through to O(eps^2),''
  JHEP\ {\bf 1011} (2010) 102
  [arXiv:1010.4478 [hep-ph]].
  %%CITATION = JHEPA,1011,102;%%

%\cite{hep-th/0404092}
\bibitem{hep-th/0404092}
  A.~V.~Kotikov, L.~N.~Lipatov, A.~I.~Onishchenko and V.~N.~Velizhanin,
  %``Three loop universal anomalous dimension of the Wilson operators in N=4 SUSY Yang-Mills model,''
  Phys.\ Lett.\ B\ {\bf 595} (2004) 521
   [Erratum-ibid.\ B\ {\bf 632} (2006) 754]
  [hep-th/0404092].
  %%CITATION = PHLTA,B595,521;%%

%\cite{Magnea:1990zb}
\bibitem{Magnea:1990zb}
  L.~Magnea and G.~F.~Sterman,
  %``Analytic continuation of the Sudakov form-factor in QCD,''
  Phys.\ Rev.\ D {\bf 42} (1990) 4222.
  %%CITATION = PHRVA,D42,4222;%%

%\cite{Korchemskaya:1992je}
\bibitem{Korchemskaya:1992je}
  I.~A.~Korchemskaya, G.~P.~Korchemsky,
  %``On lightlike Wilson loops,''
  Phys.\ Lett.\  {\bf B287 } (1992)  169-175.

%\cite{Grisaru:1982zh}
\bibitem{Grisaru:1982zh}
  M.~T.~Grisaru, W.~Siegel,
  %``Supergraphity. 2. Manifestly Covariant Rules and Higher Loop Finiteness,''
  Nucl.\ Phys.\  {\bf B201 } (1982)  292.
  
%\cite{Marcus:1984ei}
\bibitem{Marcus:1984ei}
  N.~Marcus, A.~Sagnotti,
  %``The Ultraviolet Behavior Of N=4 Yang-mills And The Power Counting Of Extended Superspace,''
  Nucl.\ Phys.\  {\bf B256 } (1985)  77.  

\end{thebibliography}
\end{document}